\documentclass[twocolumn,aps,prb,showpacs]{revtex4}
\usepackage{graphicx}
\usepackage{dcolumn}
\usepackage{amssymb}
\begin{document}

\title{Critical behavior of the Coulomb-glass model in the zero-disorder
limit: \linebreak Ising universality in a system with long-range 
interactions}

\author{A.~M\"obius\footnote{e-mail: a.moebius@ifw-dresden.de} 
and U.K.~R\"o{\ss}ler}

\affiliation{Leibniz Institute for Solid State and Materials Research 
      IFW Dresden, PF 27 01 16, D-01171 Dresden, Germany}
      
\date{\today}

\begin{abstract}     
The ordering of charges on half-filled hypercubic lattices is 
investigated numerically, where electroneutrality is ensured by 
background charges. This system is equivalent to the $s = 1/2$ Ising 
lattice model with antiferromagnetic $1/r$ interaction. The temperature 
dependences of specific heat, mean staggered occupation, and of a 
generalized susceptibility indicate continuous order-disorder phase 
transitions at finite temperatures in two- and three-dimensional 
systems. In contrast, the susceptibility of the one-dimensional system 
exhibits singular behavior at vanishing temperature. For the two- and 
three-dimensional cases, the critical exponents are obtained by means of 
a finite-size scaling analysis. Their values are consistent with those 
of the Ising model with short-range interaction, and they imply that the 
studied model cannot belong to any other known universality class. 
Samples of up to 1400, $112^2$, and $22^3$ sites are considered for 
dimensions 1 to 3, respectively.
\end{abstract} 


\maketitle

\section{Introduction}

The long range of electrostatic correlations underlies important 
physical effects such as screening, charge renormalization, 
charge orders and instabilities of plasmas.\cite{Levin02} In Coulomb 
glasses, disordered systems of localized charges, electrostatic 
correlations induce the Coulomb gap in the single-particle energy 
spectrum.\cite{OTCD01} The question under which conditions these glasses 
exhibit genuine glass transitions has been under controversial debate 
for a long 
time.\cite{V93,GY93,VS94,Detal00,Ovetal04,PaDo,MuePa,MaKu,GoePa,Suetal}

Lattices partially occupied by charged particles have been studied as 
models for ionic fluids, where several phases were identified and an 
Ising-type liquid-gas critical point was 
observed.\cite{DS99,Letal02,Brogetal,Ciach.Stell.03,Ciach.Stell.04,Ciach} 
In particular, by means of numerical simulation and finite-size scaling,
Luijten {\it et al.}\ could clearly rule out this point to belong to any
of several alternative universality classes.\cite{Letal02} For a recent 
review on this field see Ref.\ \onlinecite{Ciach.Stell.05}. In case of  
full occupancy, such a lattice corresponds to an ionic crystal and 
provides the zero-disorder limit of the Coulomb-glass problem. 

Although the ordering of ionic crystals with decreasing temperature 
has been studied by several groups, important questions are still open.
Analytical\cite{Ciach.Stell.03,Ciach.Stell.04,Ciach} and simulation 
results \cite{Alma.Enci,MoeRoe.phy,MoeRoe.arc,Ovetal04} suggest that, 
for the case of pure Coulomb interaction, staggered (antiferromagnetic) 
ordering in three-dimensional simple-cubic lattices starts with a 
continuous phase transition. According to the hierarchical reference 
theory study by Brognara {\it et al.}\cite{Brogetal} and the 
renormalization-group investigation by Ciach,\cite{Ciach} this 
transition should belong to the short-range Ising universality class in 
spite of the long-range interaction. This hypothesis is suggested by the
transition being the end of a line of Ising transitions in the lattice 
restricted primitive model, which is reached for complete filling. 

However, we remind that, in the case of Ising models with ferromagnetic 
long-range interactions, the phase transition depends on dimensionality
and decay exponent of the interactions. The critical indices can vary 
from mean-field behavior to that of the short-range Ising universality 
class via intermediate values.\cite{Fi.etal.72,LB02} 

The case of antiferromagnetic long-range interactions is more subtle due 
to their inherent frustration. From the corresponding field-theoretic 
approach for a staggered order, one would expect that the long-distance 
behavior of the interactions should renormalize to short-range 
couplings.\cite{Ciach.Stell.03} However, this argument assumes validity 
of a perturbative expansion around the ground-state. Thus, one discounts 
part of the couplings between charge or higher-order multipole 
fluctuations that are associated with the defects of the ordered state.
In the case of Coulomb-glass systems, the interplay of such fluctuations 
and quenched disorder is believed to underly glassy properties. Recently, 
three different analytical approaches for the screening problem in such 
disordered electronic systems have been developed, where two of them 
contain the non-disordered case as natural limit.\cite{PaDo,MuePa,MaKu}
In this context, precise numerical experiments on the phase transition 
of lattice models without disorder are called for.

The corroboration of the analytical results on the ordering of ionic 
crystals by numerical studies is very difficult due to the long-range 
interaction: Studying samples of up to $12^3$ sites, Almarza and Enciso 
observed their data for simple-cubic lattices to be consistent with the 
assumption of short-range Ising universality.\cite{Alma.Enci} However, 
they could not determine the critical exponents directly because the 
samples were still too small. A considerable progress was reached in 
Refs.\ \onlinecite{MoeRoe.phy} and \onlinecite{MoeRoe.arc}, where 
samples of up to $18^3$ sites were considered, so that critical 
exponents could be obtained by finite-size scaling. The exponent values 
were found to be very close to the data for the short-range Ising model, 
but this work was presented only in a short form. Moreover, further 
enlarging of the sample size and diminishing of the error bars seemed 
desirable. Finally, Overlin {\it et al.}\ studied the influence of 
positional disorder at samples of up to $8^3$ sites and determined the 
critical exponent of the localization length.\cite{Ovetal04} For the 
limiting case of vanishing disorder, their numerical result was 
consistent with both the short-range Ising and mean-field universality 
classes.

The effects of unscreened interactions and frustration may become
more prominent in dimensions $d=1$ and 2. However, only a few numerical
studies have been devoted to the behavior of such systems: In a first 
attempt, D\'\i az-S\'anchez {\it et al.}\ studied samples of up to 26 
and $14^2$ sites for $d = 1$ and 2, respectively, by means of a 
spin-glass approach.\cite{Detal00} They concluded that a phase 
transition at finite $T$ exists only for $d = 2$, but not for $d = 1$. 
The behavior of considerably larger samples was simulated in Refs.\ 
\onlinecite{MoeRoe.phy} and \onlinecite{MoeRoe.arc}, where a 
continuous phase transition at finite $T$ was obtained for $d = 2$, but 
not for $d = 1$. In the former case, according to the obtained critical 
exponents, short-range Ising universality seems likely.  Moreover, Luo 
{\it et al.}\ considered two-dimensional systems with logarithmic 
interaction, corresponding to the interaction of homogeneously charged 
lines.\cite{Luoetal} They obtained values of the critical exponents of 
correlation length and order parameter which clearly differ from the 
short-range Ising values.

The aim of the present work is twofold: On the one hand, for $d = 3$, 
we numerically investigate the ordering in simple-cubic systems to 
check the analytical theories mentioned above. The critical exponents of 
correlation length, order parameter, specific heat, and generalized 
susceptibility are obtained by finite-size scaling. On the 
other hand, we extend these studies to the cases $d = 1$ and 2. For a 
preliminary and less detailed version of this work, which was still 
restricted to the investigation of smaller systems, see Ref.\ 
\onlinecite{MoeRoe.arc}.

This paper is organized as follows. In Sec.\ II, we introduce the model 
Hamiltonian including the used boundary conditions, and we comment on 
the applied numerical procedures. Section III presents a qualitative 
overview of the simulation results, whereas Sec.\ IV is devoted to a 
quantitative evaluation of the data sets by means of finite-size 
scaling. Finally, in Sec.\ V, the obtained critical exponent values are 
discussed and compared with previous work.

\section{Model and numerical approach}

We numerically investigate the nature of the order-disorder transition 
in a system of charges considering the minimal model: simple-hypercubic 
lattices are half-filled with particles interacting via the Coulomb 
potential where background charges -1/2 are attached at each site for 
neutrality,
\begin{equation}
H = \frac{1}{2} \sum_{i \neq j} f_{ij} (n_i - 1/2)\,(n_j - 1/2)
\end{equation}
with 
\begin{equation}
f_{ij} = 1 / \left|{\bf r}_i-{\bf r}_j\right|\,\,.
\end{equation}
Here $n_\alpha \in \{0,1\}$ denote the occupation numbers 
of states localized at sites ${\bf r}_i$ within a $d$-dimensional 
hypercube of size $L^d$. Elementary charge, lattice spacing, dielectric 
and Boltzmann constants are all taken to be 1, so that the temperature
$T$ is dimensionless. Due to particle-hole symmetry, canonical and grand 
canonical Hamiltonians equal each other in this case. When substituting 
$s_i$ for $n_i - 1/2$, Eqs.\ (1) and (2) take the form of the $s = 1/2$ 
Ising lattice model with Coulomb interaction of antiferromagnetic
character.

For reducing finite-size effects, we impose periodic boundary conditions
for $d = 1$ and 2, modifying $f_{ij}$ according to the minimum image 
convention.\cite{Metro53} For $d = 3$, the same approach would give rise 
to an unphysical feature: The groundstate would be a layered arrangement 
of charges instead of the expected NaCl structure in case $L$ is a 
multiple of four.\cite{Metal01} Therefore, similar to Ref.\ 
\onlinecite{Napoetal}, we consider the sample to be surrounded by 26 
equally occupied cubes in this case,\cite{boucon}
\begin{equation}
f_{ij} = \frac{1}{\left|{\bf r}_i-{\bf r}_j\right|} + 
\sum_{k=1}^{26} (\frac{1}{\left|{\bf r}_i-{\bf r}_j-{\bf R}_k\right|} - 
\frac{1}{\left|{\bf R}_k\right|})\,\,.
\end{equation}
Here, ${\bf R}_k$ denotes the shift of the neighboring cube $k$ with 
respect to the central cube. Similar to the minimum image convention, 
the correction terms in Eq.\ (3) efficiently reduce the largest 
finite-size effect, namely the difference between the surroundings of 
sites close to the center of the cube and of sites close to its surface. 
Compared to the implementation of periodic boundary conditions combined 
with an Ewald summation, our method has the advantage that it does not
introduce the artificial long-range correlations arising from the series 
of periodic images of the cell. However, in the limit 
$L \rightarrow \infty$, both approaches should yield the same results, 
see the comparison with Ref.\ \onlinecite{Ovetal04} in Sec.\ IV.

To obtain ensemble averages of various observables, we follow the
Metropolis approach and substitute temporal for ensemble 
averaging.\cite{Metro53} Since only equilibrium properties are of 
interest here, we are free to choose the dynamics so that the 
simulation effort is minimized. A cluster Monte Carlo algorithm seems 
not to be available for the antiferromagnetic Coulomb interaction 
because of frustration. Thus we select the system modifications to be 
taken into account by hand: We include one-electron exchange with the 
``surroundings'', one-electron hops over distances below a certain 
bound, and two-electron hops changing the occupation of four neighboring 
sites. The maximum permitted distance of one-electron hops is enlarged 
when $T$ is lowered.

At high $T$, we use the original Metropolis method.\cite{Metro53} But 
at low $T$, we take advantage of the hybrid procedure proposed in Ref.\
\onlinecite{MT97}, which much accelerates the computations: Similar to
the n-fold way algorithm,\cite{Bort.Kalo} we deterministically calculate 
the rates of the transitions to all multiparticle states, which are 
accessible from the current state by means of a single system 
modification. Thus the dwelling time at the current state can be 
determined directly, and only one Monte Carlo step is needed per system 
modification. Moreover, the hybrid procedure connects the deterministic 
evaluation of weighted sums over all states within a low-energy subset 
of the configuration space with Monte Carlo sampling of the 
complementary high-energy subset. These two ideas had proved to be very 
efficient in studying the specific heat of the Coulomb glass at low 
temperatures.\cite{Metal01} 

\begin{figure}
\includegraphics[width=0.75\linewidth]{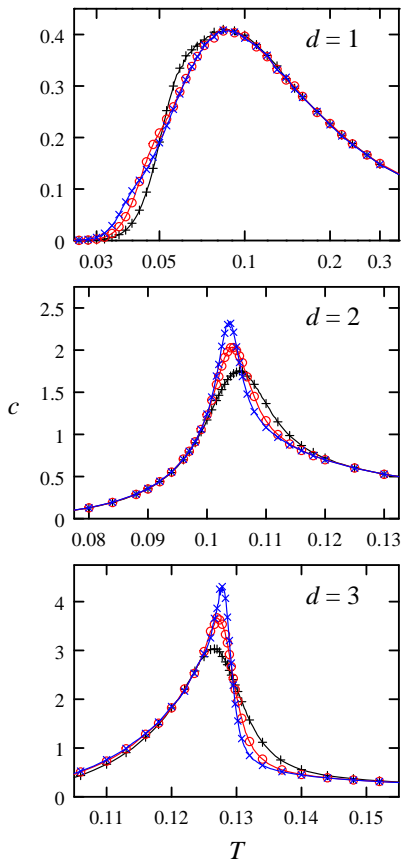}
\caption{(Color online) Temperature dependences of the specific heat, 
$c(T)$, for dimensions $d = 1$ to 3 as obtained from simulations of 
samples of $L^d$ sites.
$d = 1$: $L = 100$ ($+$), 280 ($\circ$), and 700 ($\times$); 
$d = 2$: $L = 20$ ($+$), 34 ($\circ$), and 58 ($\times$); 
$d = 3$: $L = 8$ ($+$), 12 ($\circ$), and 18 ($\times$).
Only a part of the data points forming the curves is marked by symbols, 
as well as in Figs.\ 2 to 4, 6, and 8. The error bars are considerably 
smaller than the symbol size.}
\end{figure}

\begin{figure}
\includegraphics[width=0.75\linewidth]{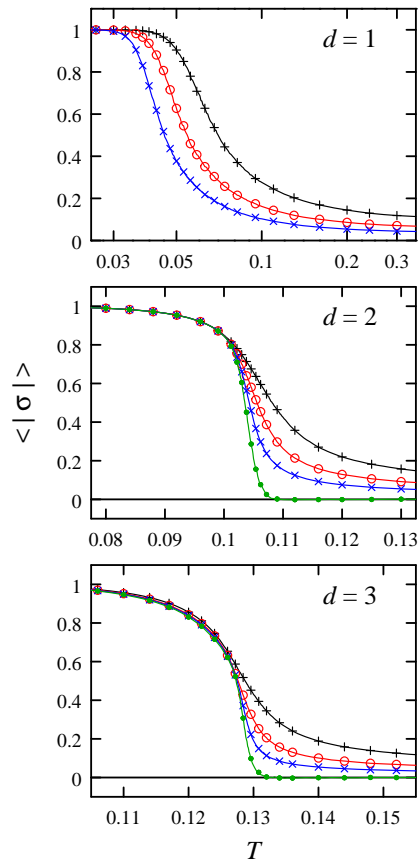}
\caption{(Color online) Temperature dependence of the average absolute 
value of the mean staggered occupation, $\langle |\sigma| \rangle(T)$, 
which is defined by Eq.\ (5) for $d = 1$ to 3. For the meaning of the 
symbols $+$, $\circ$, and $\times$ see caption of Fig.\ 1; $\bullet$ 
marks the extrapolation $L \rightarrow \infty$ explained in the text.}
\end{figure}

For an efficient error control, we decompose the simulation time 
considering 100 intervals with integration time $\tau$ instead of one
interval of length $100\,\tau$. In detail, we increase $\tau$
step by step performing 50 runs for every $\tau$ value. In each of these 
runs, after starting randomly from one multiparticle state out of a set 
of previously tabulated low-energy states, the sample is first 
equilibrated during a time interval $\tau/3$. Then the evolution over 
two successive time intervals $\tau$ is emulated. The obtained ensembles 
of 100 measuring data for each observable are used to estimate mean 
values and their statistical uncertainties, and to check equilibration. 
Based on these results, it is decided, whether the iteration process can 
be stopped, or whether $\tau$ has to be increased further.

\section{Qualitative results}

We now turn to the qualitative behavior of specific heat, order 
parameter, and susceptibility. The specific heat $c$ was obtained from 
energy fluctuations utilizing 
\begin{equation}
c = (\langle H^2 \rangle - \langle H \rangle^2) / (T^2\ L^d) \,\,,
\end{equation}
see e.g.\ Ref.\ \onlinecite{NB}. Figure 1 shows its $T$ and $L$ 
dependences: For $d = 2$ and 3, sharp peaks of increasing height evolve 
within a small $T$ region as $L$ grows. Away from the peaks, within the 
$T$ intervals presented, $c$ is almost independent of $L$. However, for 
$d = 1$, there are only broad rounded peaks with $L$-independent height 
-- a logarithmic $T$ scale is used for $d = 1$, in contrast to the 
linear scales for $d = 2$ and 3, which display far smaller $T$ 
intervals. For $d = 1$, finite-size effects are restricted to low $T$ 
where the reliability bound decreases as $L$ grows, compare Ref.\ 
\onlinecite{Metal01}. 

\begin{figure}
\includegraphics[width=0.75\linewidth]{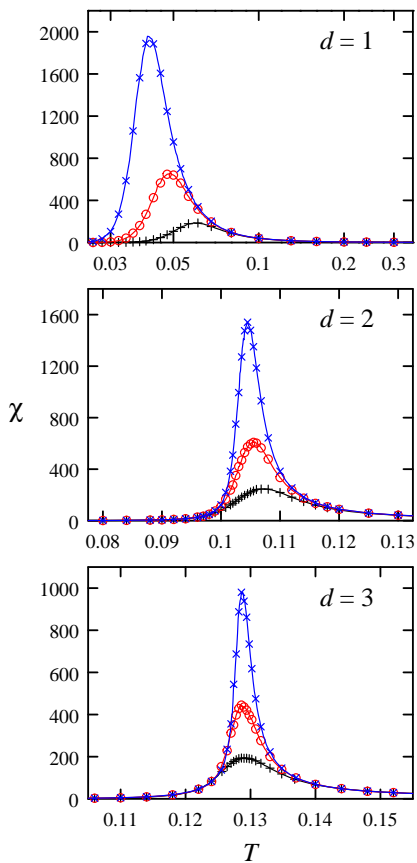}
\caption{(Color online) Temperature dependence of the susceptibility, 
$\chi(T)$, which is related to the mean staggered occupation $\sigma$ by 
Eq.\ (6), for $d = 1$ to 3. For the meaning of the symbols see caption 
of Fig.\ 1.}
\end{figure}

\begin{figure}
\includegraphics[width=0.75\linewidth]{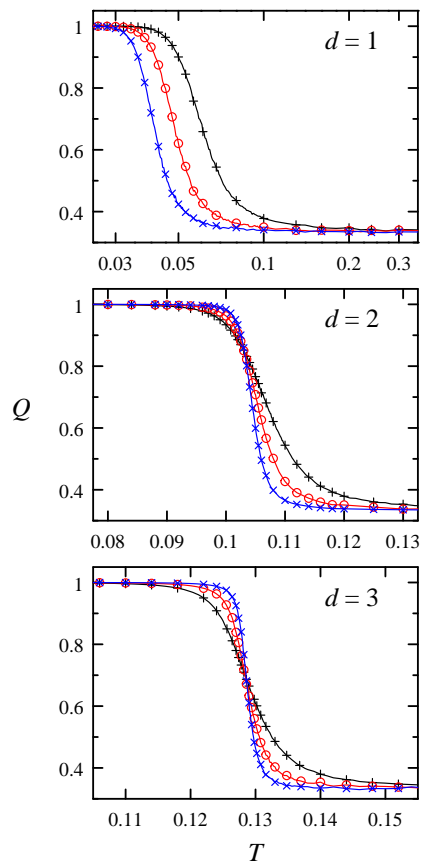}
\caption{(Color online) Temperature dependence of the Binder parameter, 
$Q(T)$, which is related to the mean staggered occupation $\sigma$ by 
Eq.\ (7), for $d = 1$ to 3. For the meaning of the symbols see caption 
of Fig.\ 1.}
\end{figure}

Analogously to an antiferromagnet, the order inherent in a charge 
arrangement $n_i$ can be characterized by the mean staggered 
occupation $\sigma$ relating to a NaCl structure.\cite{V93} For $d = 3$, 
\begin{equation}
\sigma = \frac{1}{L^d} \sum_i (-1)^{x_i + y_i + z_i} (2 n_i - 1)
\end{equation}
where $x_i$, $y_i$, and $z_i$ denote the (integer) components of 
${\bf r}_i$. Thus we consider the ensemble average of the absolute 
value of the mean staggered occupation $\langle |\sigma| \rangle$ as 
order parameter.  $T$ and $L$ dependences of $\langle |\sigma| \rangle$ 
are shown in Fig.\ 2. For $d = 1$, a rapid decrease of 
$\langle |\sigma| \rangle$ with increasing $T$ occurs already clearly 
below the temperature of maximum $c$ (same $T$ scales in Figs.\ 1 and 
2). This marked decrease shifts to lower $T$ with increasing $L$. For 
$d = 2$ and 3, a qualitatively different behavior is found: 
$\langle |\sigma| \rangle$ decreases rapidly just in that $T$ region 
where the peak of $c(T)$ evolves, and the $T$ interval of rapidly 
diminishing $\langle |\sigma| \rangle$ shrinks as $L$ rises. In these
cases, the extrapolation 
$L \rightarrow \infty$ by means of $\langle |\sigma| \rangle(T,L) =
\langle |\sigma| \rangle(T,\infty) + A(T)/L^{d/2}$ seems natural. 
Based on the $\langle |\sigma| \rangle(T,L)$ data for $L =34$ and $58$
in case $d = 2$, and for $L = 12$ and 18 in case $d = 3$, it yields 
almost sharp transitions. However, this extrapolation is of limited 
accuracy in the immediate vicinity of the transition.

The generalized susceptibility $\chi$, which is related to the 
response of $\langle |\sigma| \rangle$ to a staggered field, is given by
\begin{equation}
\chi = L^d ( \langle \sigma^2 \rangle -
\langle |\sigma| \rangle^2) / T \ ,
\end{equation}
see Ref.\ \onlinecite{BH}. Figure 3 shows $T$ and $L$ dependences of 
$\chi$. On the one hand, for $d = 1$, a broad peak of $\chi(T)$ evolves 
with increasing $L$ where $T_{\rm{max}}$, the temperature of maximum 
$\chi$, decreases. On the other hand, for $d = 2$ and 3, as $L$ rises, a 
narrow peak grows in just that $T$ region where $c(T,L)$ has such a 
feature.

Hence, according to the described behavior of $c(T,L)$, 
$\langle |\sigma| \rangle(T,L)$, and $\chi(T,L)$, a phase transition 
likely occurs for $d = 2$ and 3, for $d = 3$ in agreement with Refs.\
\onlinecite{Ovetal04,DS99,Letal02}. However, for $d = 1$, in spite of 
the long-range interaction, there seems to be no phase transition at 
finite $T$.

This conclusion is confirmed by the behavior of the Binder parameter, 
the fourth-moment ratio,
\begin{equation}
Q = \langle \sigma^2 \rangle^2 / \langle \sigma^4 \rangle\, ,
\end{equation}
which is shown in Fig.\ 4.\cite{Letal02} This quantity is directly 
derived from the Binder cumulant 
$1 - \langle \sigma^4 \rangle / (3 \langle \sigma^2 \rangle^2)$ and 
exhibits similar features: In the ordered phase, where $\sigma$ has a 
bimodal probability distribution, $Q = 1$, whereas, in case of a 
normal distribution of $\sigma$ in the disordered phase, $Q = 1/3$. 
A common intersection point of the $Q(T)$ curves for various system 
sizes indicates a phase transition. According to Fig.\ 4, such points
seem to exist for $d = 2$ and 3, but not for $d = 1$.

\section{Quantitative analysis by means of finite-size scaling}

Consider first the one-dimensional case: Careful inspection of Fig.\ 4
leads to the conclusion that, although there seems to be no common
intersection point of the $Q(T)$ curves, simple scaling 
$Q(T,L) = Q(T/L^p)$ is unlikely because of the $L$ dependence of the 
slope in this graph. Thus, to perform a detailed analysis, we 
numerically solved $Q(T_A,L) = A$ for $A = 0.45$, 0.60, 0.75, and 0.90. 
The resulting $T_A(L)$ are presented in a $1/T_A$ versus $\mbox{ln} L$ 
plot in Fig.\ 5. 

Figure 5 shows that, with high precision, $1/T_A$ is a linear function 
of $\mbox{ln} L$ for all considered $A$ values. (The tiny deviation at 
$L = 70$ arises from a small systematic shift of the $Q(T,L)$ curves for 
$L = 4 n$, where $n$ is integer, in comparison to the curves with 
$L = 4 n + 2$. This shift vanishes as $n \rightarrow \infty$.) Moreover, 
it is remarkable that the slope of the linear function is independent of 
$A$ within the accuracy of the simulations. Finally, with increasing 
$L$, the $L$ dependence of the maximum temperature $T_{\rm max}$ of the 
susceptibility seems to tend to the same behavior as $T_A(L)$. 

Due to the linear dependences in Fig.\ 5, $T_A(L)$ and $T_{\rm max}(L)$ 
likely vanish as $L \rightarrow \infty$. This is confirmed by the 
parallelism of the regression lines, a second argument against the
intersection of $Q(T,L = {\rm const})$ curves at finite $T$. As a 
consequence of these two observations, $Q(T,L)$ is expected to depend 
only on the composed quantity $z = T_0/T - \mbox{ln}L$ with 
$T_0 = 0.2484(8)$, where the uncertainty denotes the $3 \sigma$ bound 
of the random deviation. This reduction is confirmed by Fig.\ 6, which 
shows the corresponding master curve made up of $Q(T,L = {\rm const})$ 
curves with $L = 40$, 100, 280, 700, and 1400.

Thus, in the limit $L = \infty$, long-range order should be destroyed by 
thermal excitations at arbitrarily small, but finite $T$ although the
considered model exhibits a long-range interaction.

One may ask whether the logarithmic relation indicates that
the antiferromagnetic Coulomb interaction is a marginal case for 
$d = 1$. Thus we performed additional simulations for the 
antiferromagnetic $1/r^{1/2}$ interaction which decays more slowly. 
These studies yielded similar results: Considering samples of up to 1000 
sites, we could not find a clue of a phase transition at finite $T$. 
However, on the one hand, $1/T_A$ rises with increasing $L$ slightly 
faster than as a linear function of $\mbox{ln} L$, and, on the other 
hand, small deviations from the parallelism of the dependences of 
$1/T_A$ on $\mbox{ln} L$ for different $A$ are present. 

\begin{figure}[t]
\includegraphics[width=0.75\linewidth]{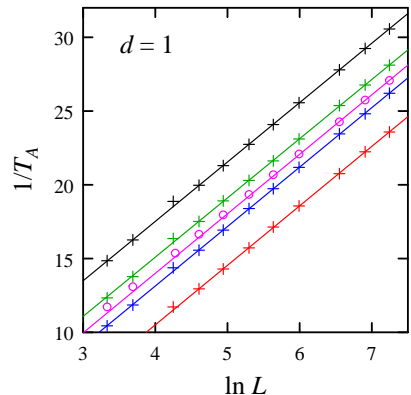}
\caption{(Color online) Size dependence of the solution $T_A$ of 
$Q(T_A,L) = A$ for $d = 1$ in a $1/T_A$ versus $\mbox{ln} L$ plot. From
top to bottom, the crosses are related to $A = 0.90$ (black), 0.75 
(green / gray), 0.60 (blue / gray), and 0.45 (red / gray). The maximum 
temperatures $T_{\rm max}$ of $\chi(T,L)$ are included as magenta (gray) 
circles for comparison. Error bars are omitted since they are 
considerably smaller than the symbol size. The straight lines correspond 
to linear regression in this representation for $L \ge 100$ in case of 
$T_A$, and for $L \ge 700$ in case of $T_{\rm max}$.}
\end{figure}

\begin{figure}
\includegraphics[width=0.75\linewidth]{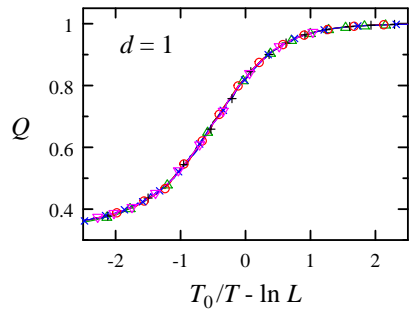}
\caption{(Color online) Check for $d = 1$ whether $Q(T,L)$ depends only 
on the composed quantity $z = T_0 / T - \mbox{ln}L$ where 
$T_0 = 0.2484$. The symbols $\bigtriangleup$  (green / gray), $+$ 
(black), $\circ$ (red / gray), $\times$ (blue / gray), and 
$\bigtriangledown$ (magenta / gray) stand for $L = 40$, 100, 280, 700, 
and 1400, respectively.}
\end{figure}

We turn now to the two- and three-dimensional cases: Here, the 
quantitative evaluation of our simulation data consists in a finite-size 
scaling analysis.\cite{Betal00,G92} However, for numerical convenience, 
we consider the quantities 
\begin{equation}
q_2 = -\ln(1 - Q)
\end{equation}
and 
\begin{equation}
q_3 = -\tan(\pi\,(1 - 1.5 \, Q)) 
\end{equation}
for $d = 2$ and 3, respectively, instead of $Q$. The $q_d(T)$ have small 
curvature in the vicinity of the transition. This behavior alleviates a 
precise interpolation.

Figure 7 shows $q_d(T)$ for various $L$ values. For $d = 2$, there 
clearly is a common intersection point of the curves for different $L$ 
at the critical temperature $T_{{\rm c},2}$. However, for $d = 3$, only 
a tendency toward such a behavior is seen although the widths of the 
$T$ intervals for both cases as well as the ranges of numbers of sites 
$L^d$ are comparable. 

The corresponding corrections to scaling can be taken into account to a 
large extent in a simple way: Following Ref.\ \onlinecite{Hase}, we 
define a size-dependent critical temperature $T_{{\rm c},d}(L)$ by the 
demand $q_d(T_{{\rm c},d}(L),L) = q_{0,d}$, where the $q_{0,d}$ are 
appropriate constants, which are fixed below. Scaling of $q_d(T)$ curves 
for different $L$ with respect to $T - T_{{\rm c},d}(L)$ yields very 
good data collapse. Figure 8 demonstrates this observation showing plots 
of $Q(T,L)$ versus $t = a_d(L) (T - T_{{\rm c},d}(L))$ based on the 
parameter values obtained below. In our study, referring to 
$T_{{\rm c},d}(L)$ instead of $T_{{\rm c},d}(\infty)$ proved to 
considerably reduce the influence of deviations from scaling on the 
values of the critical exponents, which are numerically obtained from 
samples of finite size.

\begin{figure}[t]
\includegraphics[width=0.75\linewidth]{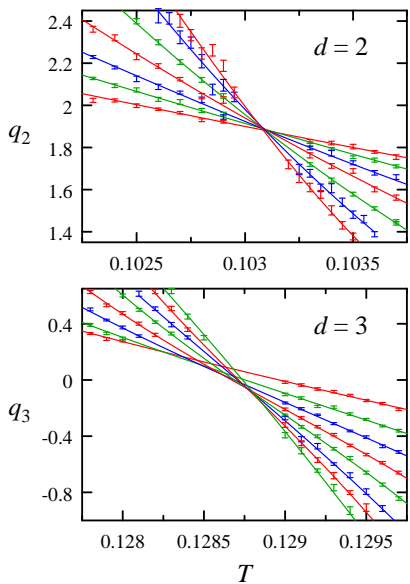}
\caption{(Color online) Temperature dependence of $q_d$, related to the 
Binder parameter $Q$ by Eqs.\ (8) and (9), in the close vicinity of the 
transition for $d = 2$ and 3. According to increasing modulus of the 
slope, the curves refer to $L = 16$, 24, 34, 48, 68, 88, and 112 for 
$d = 2$, and to $L = 8$, 10, 12, 14, 16, 18, 20 and 22 for $d = 3$. The
error bars denote the $1 \sigma$ region. For clarity, data points in the 
intersection regions are omitted.}
\end{figure}

\begin{figure}
\includegraphics[width=0.75\linewidth]{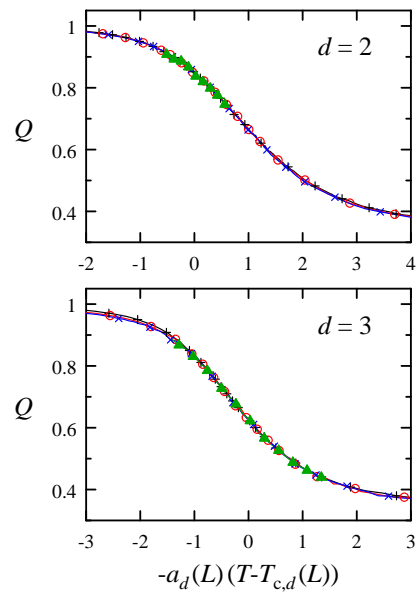}
\caption{(Color online) Scaling check for $Q(T,L)$: The adjustment of 
the parameters $a_d(L)$ and $T_{{\rm c},d}(L)$ is described in the text.
Scaling is indicated by the agreement of the nonlinear contributions to
$Q(T,L = {\rm const})$ for various $L$ in these plots. For the meaning 
of the symbols $+$, $\circ$, and $\times$ see caption of Fig.\ 1; 
$\blacktriangle$ (green / gray) refers to $L = 112$ and 22 for $d = 2$ 
and 3, respectively.}
\end{figure}

Figure 8 simultaneously shows that $Q(t)$ has a substantial curvature 
within the crucial region -- our finite-size scaling analysis is based 
on the $Q$ intervals $[0.75,0.91]$ and $[0.45,0.80]$ for $d = 2$ and 3, 
respectively.  However, to reach a high accuracy of the critical 
exponents, very precise $a_d(L)$ values are needed so that a broad $t$ 
range has to be taken into account. We approach this nonlinearity 
problem by considering $q_d(t)$ instead of $Q$ on the one hand, and by 
approximating $q_d(t)$ by polynomials of third degree, 
$q_{0,d} + t + b_d t^2 + c_d t^3$, on the other hand.

We evaluated our $q_d(T,L)$ data by a series of regression studies of 
the $T$ and $L$ dependences: First, the only weakly $L$-dependent 
parameters $b_d$ and $c_d$ of the polynomial ansatz were adjusted, after 
this the $a_d(L)$ and $T_{{\rm c},d}(L)$ values were determined. 

Both for the cases $d = 2$ and 3, the relations $a_d(L)$ only weakly 
deviate from power laws, $a_d(L) \propto L^p$, so that they are not 
graphically depicted here.  Based on the $a_d(L)$ data, the value of the 
critical exponent of the correlation length, $\nu = 1/p$, can be 
obtained in two ways. First, it can be directly calculated by numerical 
differentiation by means of the midpoint formula utilizing
$\nu = (\mbox{d} \mbox{ln} | a_d | / \mbox{d} \mbox{ln} L)^{-1}$.
The advantage of this approach is a meaningful control of convergence 
with increasing $L$. Figure 9 shows such results from the consideration 
of pairs of sixth-next and next-nearest neighbors in the sequence of 
sample sizes for $d = 2$ and 3, respectively. Second, $\nu$ can be 
determined by means of power-law fits taking into account various $L$ 
intervals. This method yields more precise estimates of the exponents.
For consistency, the mean-square deviation of these fits must be 
understandable as resulting from random errors alone. Table I presents 
the most precise results for $\nu$, which were obtained from the fits 
safely fulfilling this requirement. Additionally, these values are 
included in Fig.\ 9. 

\begin{figure}
\includegraphics[width=0.75\linewidth]{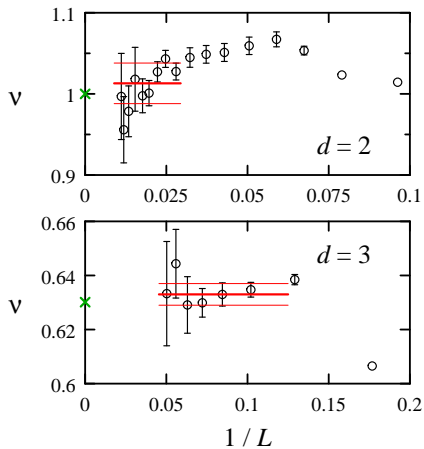}
\caption{(Color online) Approximations of the critical exponent $\nu$ 
versus sample size. The symbols $\circ$ denote values which were 
obtained by numerical differentiation, see text. The corresponding
error bars denote the $1 \sigma$ regions. The thick red (gray) lines
represent results of power-law fits, where their extension visualizes 
the evaluated $L$ interval. The thin red (gray) lines give upper and
lower bounds of the corresponding $3 \sigma$ intervals. For comparison,
the green (gray) symbols $\times$ mark the known values of the Ising 
model with short-range interactions for $L = \infty$, see Tab.\ I.}
\end{figure}

\begin{figure}
\includegraphics[width=0.75\linewidth]{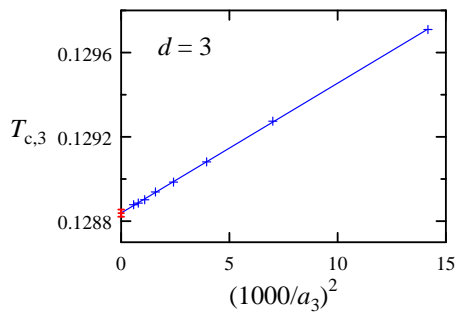}
\caption{(Color online) Relation between $a_3(L)$ and $T_{{\rm c},3}(L)$ 
for $q_{0,3} = -0.200$. With increasing $a_3$, the points refer to 
$L = 8$, 10, 12, 14, 16, 18, 20, and 22.  The error bar represents our 
extrapolation $L \rightarrow \infty$, where the $3 \sigma$ interval is 
marked.}
\end{figure}

This graph implies several conclusions: The $\nu$ values converge 
rapidly with increasing sample size so that, for $d = 3$, already the 
study of samples with $L \approx 12$ yields $\nu$ values, which are very 
close to the results for $L \approx 20$. Thus the good agreement between
our result of a power-law fit, 0.633(4), and  the known Ising value 
0.630, suggests that the considered model may belong to the Ising 
universality class. This hypothesis is supported by our result 0.633(4) 
clearly differing from the mean-field value 0.5, from the Heisenberg 
value 0.71, and from exponents of the alternative ``nearby'' models XY, 
$\nu = 0.670$, and self-avoiding walks, $\nu = 0.588$, compare Ref.\ 
\onlinecite{Letal02}. Also for $d = 2$, our $\nu$ value supports Ising
universality and clearly excludes mean-field behavior. Finally, 
according to Fig.\ 9, it is not surprising that, for $d = 3$, Overlin 
{\it et al.}\ got the somewhat lower value $0.55 \pm 0.1$ from 
simulations for $L = 4$, 6, and 8.\cite{Ovetal04} 

In obtaining $T_{{\rm c},d}(L)$ from 
$q_d(T_{{\rm c},d}(L),L) = q_{0,d}$, a deviation $\delta$ of $q_{0,d}$ 
from the $L \rightarrow \infty$ limit of the solution of  
$q_d(T,2 L) = q_d(T,L)$ gives rise to a contribution 
$\delta / a_d(L)$ to $T_{{\rm c},d}(L)$. We fix $q_{0,d}$ by the demand 
that this term vanishes: $q_{0,2} = 1.892(8)$ and 
$q_{0,3} = -0.200(19)$, corresponding to the $Q$ values 0.8492(12) and 
0.625(4), respectively. For $d = 2$, the critical $Q$ value slightly,
but significantly deviates from the analytical result for the 
$L = \infty$ limit of the short-range Ising model with periodic boundary 
conditions 0.856216.\cite{KaBl} But this is not a strong counterargument
to the considered model belonging to the Ising universality class, as it
is suggested by the value of $\nu$: It is known that the critical $Q$ 
value is a very sensitive quantity which depends on the boundary 
conditions, and that ``universality of the critical cumulant holds in a 
rather restricted sense, when compared to universality of critical 
exponents''. \cite{Se06} Nevertheless, for $d = 3$, our critical $Q$ 
value perfectly agrees with the value for the Ising model with 
short-range interaction and periodic boundary conditions, 
0.6233(4).\cite{Bletal95} This supports the hypothetical Ising 
criticality infered from the value of $\nu$.

\begin{table}
\caption{Finite-size scaling results for the critical exponents of 
specific heat, mean staggered occupation, susceptibility, and 
correlation length, $\alpha$, $\beta$, $\gamma$, and $\nu$, 
respectively. To retain numerical precision, we mostly present exponent 
ratios instead of the exponents themselves. The data were obtained by 
two alternative methods, either directly (marked as d) by power-law 
fits (modified by a constant term in the case of $\alpha / \nu$), or 
indirectly (marked as i) via Widom and hyperscaling relations from power 
law fits yielding other exponents or exponent ratios, see text. Values 
for the Ising model with short-range interaction 
{\protect \cite{G92,Campo}} are included for comparison. Parentheses and 
brackets give $3 \sigma$ random errors and total errors, respectively, 
referring to the last given digit of the value.}
\begin{ruledtabular}
\begin{tabular}{cccccc}
Quantity & $d$ & $L$ region & method & Coulomb&s.-r.~Ising \\
\hline
$\alpha / \nu$&2&28 -- 112&d&-0.02(4)&0 (ln)\\
$\alpha / \nu$&2&34 -- 112&i&-0.03(5)&0 (ln)\\
$\beta / \nu$&2&48 -- 112&d&0.1318(21)&1/8\\
$\beta / \nu$&2&48 -- 112&i&0.129(8)&1/8\\
$\gamma / \nu$&2&48 -- 112&d&1.742(15)&7/4\\
$\nu$&2&34 -- 112&d&1.013(25)&1\\
$\alpha / \nu$&3&10 -- 22&d&0.09(9)&0.1740[8]\\
$\alpha / \nu$&3&8 -- 22&i&0.158(21)&0.1740[8]\\
$\beta / \nu$&3&14 -- 22&d&0.506(7)&0.51820[8]\\
$\beta / \nu$&3&14 -- 22&i&0.514(5)&0.51820[8]\\
$\gamma / \nu$&3&14 -- 22&d&1.973(10)&1.96361[15]\\
$\nu$&3&8 -- 22&d&0.633(4)&0.63012[16]\\
\end{tabular}
\end{ruledtabular}
\end{table}

The remaining higher-order corrections in $T_{{\rm c},d}(L)$ originate 
from imperfection of finite-size scaling.  Comparing several empirical 
approximations, we observed that, over a wide $L$ range, they are almost 
proportional to $a_d(L)^{-2}$, see Fig.\ 10. Corresponding 
extrapolations yield the following values of $T_{{\rm c},d}(\infty)$: 
$0.103082(9)$ and $0.128838(17)$ for $d = 2$ and 3, respectively. The 
confidence intervals include the $3 \sigma$ random errors and cautious 
estimates for the systematic uncertainty of the extrapolation 
$L \rightarrow \infty$.

Our $T_{{\rm c},3}$ value is consistent with the result 
$0.128 \pm 0.005$ in Ref.\ \onlinecite{Ovetal04}, which was obtained for 
considerably smaller samples. This coincidence, together with the 
approximate agreement of the values of $\nu$, confirms an assumption 
from Sec.\ II. It shows that, within the accuracy of the simulation, our 
treatment of the surroundings of the sample yields the same results as 
the Ewald summation performed in Ref.\ \onlinecite{Ovetal04}. Moreover,
comparing with analytical theories, we mention that the nonlinear 
screening theory by Pankov and Dobrosavljevi\'c underestimates our 
``exact'' $T_{{\rm c},3}$ value by $26 \%$,\cite{PaDo} whereas a study 
by Malik and Kumar, which utilizes the replica method, overestimates 
$T_{{\rm c},3}$ by roughly a factor 1.5 when using the random-phase 
approximation, and by more than a factor 3 for the Hartree 
approximation.\cite{MaKu}

\begin{figure}
\includegraphics[width=0.75\linewidth]{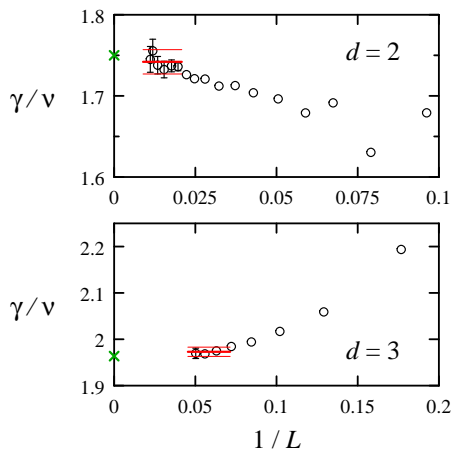}
\caption{(Color online) Approximations of the critical exponent ratios
$\gamma / \nu$ versus sample size. For the meaning of the symbols see
caption of Fig.\ 9.}
\end{figure}

\begin{figure}
\includegraphics[width=0.75\linewidth]{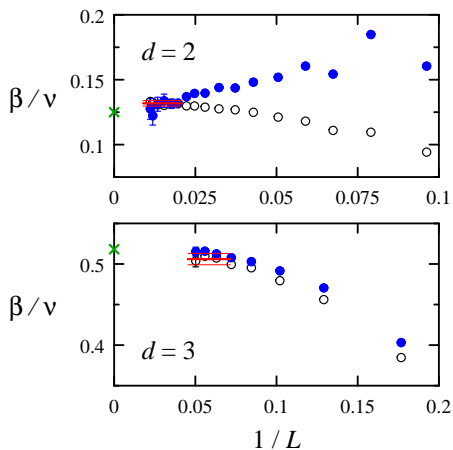}
\caption{(Color online) Approximations of the critical exponent ratios
$\beta / \nu$ versus sample size. The meaning  of most symbols is 
defined in the caption of Fig.\ 9. Here, additionally, blue (gray) 
symbols $\bullet$ denote values, which were obtained by means of the
Widom and hyperscaling relations from $\gamma / \nu$ data in Fig.\ 11, 
marked by $\circ$ there.}
\end{figure}

The analysis of $\chi(T,L)$, $\langle |\sigma| \rangle(T,L)$, and 
$c(T,L)$ was performed similarly to the evaluation of $q_d(T,L)$: We 
considered $\ln \chi$, $\ln \langle |\sigma| \rangle$, and $\ln c$ as 
functions of $t$ and $L$.  For not too large $|t|$, as 
$L \rightarrow \infty$, scaling implies that each of these quantities 
is decomposable into a sum of two functions depending only on $t$ and 
$L$, respectively. However, for the $L$ regions considered here, this 
hypothesis proved to be well fulfilled only for $\ln \chi$. In the cases 
of $\ln c$ and $\ln \langle |\sigma| \rangle$, there is a clear tendency 
toward such a behavior, but small deviations cannot be neglected. Thus 
we approximated $\ln \chi$, $\ln \langle |\sigma| \rangle$, and $\ln c$ 
by polynomials in $t$ of third order, taking advantage of universalities 
in the coefficients as far as possible. This regression provides precise 
values for the observables at $t = 0$. Simultaneously, we obtained the 
confidence intervals taking into account the uncertainties in the 
individual measurements of the observables and in the $T_{{\rm c},d}(L)$ 
values.

\begin{figure}
\includegraphics[width=0.75\linewidth]{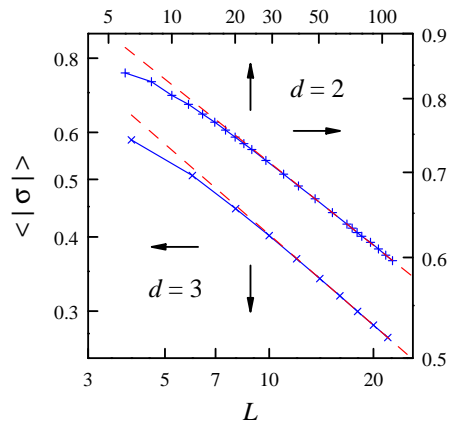}
\caption{(Color online) Size dependence of the value of 
$\langle |\sigma| \rangle$ at the critical temperature 
$T_{{\rm c},d}(L)$ defined in the text. The error bars are considerably 
smaller than the symbol size, thus they are omitted. The dashed lines 
represent the fits given in Tab.\ I.}
\end{figure}

The interpolated $\chi(T_{{\rm c},d}(L),L)$ and 
$\langle |\sigma| \rangle(T_{{\rm c},d}(L),L)$ were analyzed by means of 
power-law fits, where proportionality to 
$L^{\gamma/\nu}$ and $L^{-\beta/\nu}$, respectively, was presumed. These 
studies were performed analogously to the determination of $\nu$. 
However, while the effective $\gamma / \nu$ converges quite rapidly with 
increasing $L$, see Fig.\ 11, the determination of the limit of 
$\beta / \nu$ as $L \rightarrow \infty$ in Fig.\ 12 is hindered by slow 
convergence: Within a fixed $L$ interval, the relative change of the 
effective $\beta$ is considerably larger than the variations of the 
effective $\nu$ and $\gamma$. The reason of this slow convergence is 
understood by inspection of Fig.\ 13. Although this graph shows 
high-quality power-law behavior above $L \approx 30$ and $L \approx 10$ 
for $d = 2$ and 3, respectively, the uncertainty of the slope is rather 
large since, due to the small relative change of 
$\langle |\sigma| \rangle$, small deviations from the power law may 
considerably shift the exponent value.

Alternatively, the value of $\beta / \nu$ can be obtained from the value
of $\gamma / \nu$ utilizing the Widom relation, 
$2 = \alpha + 2 \beta + \gamma$, and the hyperscaling relation, 
$2 - \alpha = d \, \nu$. These equations imply 
$\beta / \nu = (d - \gamma / \nu) / 2$. Corresponding results are 
included in Fig.\ 12, as well as in Tab.\ I, additionally to the results
of the power-law fits. 

Table I shows that our $\beta / \nu$ data are very close to the exponent 
ratios for the short-range Ising model, where the results of the 
indirect approach deviate somewhat less from the Ising values than the 
data obtained directly by means of power-law fits.

\begin{figure}
\includegraphics[width=0.75\linewidth]{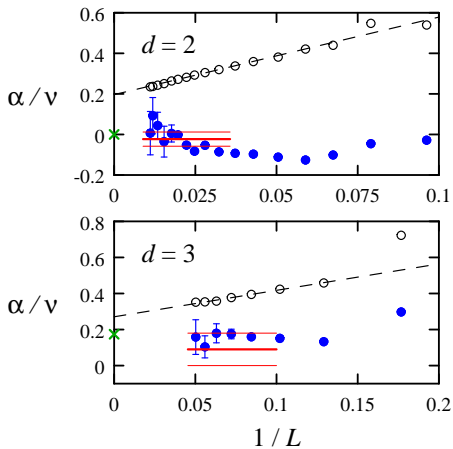}
\caption{(Color online) Approximations of the critical exponent ratios
$\alpha / \nu$ versus sample size. The meaning  of most symbols is 
defined in the caption of Fig.\ 9. However, in this case, the power-law 
fit is modified by a background constant, see text. Additionally, blue 
(gray) symbols $\bullet$ denote values, which were obtained by means of 
the hyperscaling relation from $\nu$ data in Fig.\ 9, marked by $\circ$ 
there. The dashed lines are included only as guide to the eye in order 
to demonstrate the considerable $L$ dependence of the exponent ratio 
$\alpha / \nu$ obtained by numerical differentiation.}
\end{figure}

Compared to the study of  $\langle |\sigma| \rangle$ and $\chi$, the
analysis of $c$ is more difficult: Exponent values obtained by means of
numerical differentiation converge only very slowly with increasing $L$, 
see Fig.\ 14, and the mean-square deviations of the power-law fits 
remain too large. Therefore we took into account a background 
contribution presuming $c(T_{{\rm c},d}(L),L) = a + b L^p$ with 
$p = \alpha/\nu$ similar to Ref.\ \onlinecite{Weetal}. Moreover, in 
order to avoid numerical problems, we approximated 
$c(T_{{\rm c},d}(L),L)$ by ${\overline a} + {\overline b} (L^p - 1) / p$ 
instead of by $a + b L^p$.  Results for $\alpha/\nu$, which were 
obtained in this way, are included in Tab.\ I and Fig.\ 14. 

Alternatively, $\alpha / \nu$ was calculated by means of the 
hyperscaling relation utilizing the $\nu$ results, which were obtained 
by numerical differentiation. These $\alpha / \nu$ values exhibit a far 
better convergence than the directly obtained $\alpha / \nu$ data, this 
is demonstrated Fig.\ 14. They agree nicely not only with the results of 
the modified power-law fits but also with the known values for 
short-range Ising universality, see Fig.\ 14 and Tab.\ I.

\section{Discussion}

Summarizing, we have presented a detailed Monte Carlo study of the 
ordering of charges on a half-filled hypercubic lattice. For 
one-dimensional systems, the order seems to be destroyed already at 
arbitrarily small but finite temperature in spite of the
long-range character of the interaction. However, for two- and 
three-dimensional systems, continuous phase transitions occur at finite
temperatures. We have determined the critical exponents not only for the 
correlation length, but also for the specific heat, for the order 
parameter staggered occupation, as well as for the related 
susceptibility. 

A survey of the exponent values, which we obtained by finite-size 
scaling, is given in Tab.\ I. These data have to be regarded as 
effective exponents. Due to the finiteness of $L$, tiny systematic 
errors are certainly present, presumably the more relevant the smaller 
the exponent value. Unfortunately, our data set is not sufficient for a 
convincing estimate of them. Nevertheless, the exponents which we 
obtained directly by means of power-law fits (modified by a background 
constant in case of the specific heat) obey the Widom relation, 
according to which the term 
$2 / \nu - (\alpha / \nu +  2 \beta / \nu + \gamma / \nu)$
has to vanish: Presuming the errors in Tab.\ I to be independent and
random, one obtains -0.01(6) and 0.08(9) for $d = 2$ and 3, 
respectively, where the errors are given as $3 \sigma$ bounds.
Simultaneously, our data satisfy the hyperscaling relation which 
implies the quantity $2 / \nu - \alpha / \nu - d$ to equal zero:
In this case, Tab.\ I yields -0.01(6) and 0.07(9) for $d = 2$ and 3, 
respectively. Together, Widom and hyperscaling relations imply that 
the expression $2 \beta / \nu + \gamma / \nu -d$ vanishes what can be 
checked with higher precision since $\alpha$ is not included here. For 
this expression, one obtains the values 0.006(16) and -0.015(17) for 
$d = 2$ and 3, respectively, from Tab.\ I. Thus our set of the critical 
exponent values satisfies all consistency criteria.

Among the critical exponents, $\nu$ exhibits the best convergence with
increasing sample size. The values given in Tab.\ I are consistent with 
the assumption that the studied phase transition belongs to short-range 
Ising universality class, in spite of the long-range Coulomb 
interaction, both for the cases $d = 2$ and 3. All other well-known 
universality classes could clearly be ruled out according to 
the values of $\nu$. Moreover, the supposed Ising universality is 
supported by the values of the critical exponents $\alpha$, $\beta$, and 
$\gamma$ being likewise consistent with the corresponding critical 
indices of the Ising model, as well as by the critical values of the
Binder parameter $Q$.

Concluding, in spite of the long-range interaction, the Coulomb system 
described by Eqs.\ (1) and (2) seems to belong to the same universality 
class as the Ising model with short-range interaction. This suggests 
that screening should be highly effective in the ordering process and 
that the lattice Coulomb-glass model might have the same critical 
properties as the random-field short-range Ising model.

\begin{acknowledgments}
We thank A.~Ciach, H.~Eschrig, M.E.~Fisher, B.~Kramer, T.~Nattermann, 
M.~Richter, M.~Schreiber,  R.H.~Swendsen, and T.\ Vojta for helpful 
discussions and literature hints.
\end{acknowledgments}


\begin{thebibliography}{99}
\bibitem{boucon} A single-particle hop from site $i$ to site $j$ in the 
  central cube, which is accompanied by corresponding hops in the 26
  neighboring cubes, modifies the total energy of all 27 cubes not only 
  by the difference of the corresponding single-particle energies. But, 
  moreover, it reduces the total energy by the attraction terms for the 
  729 pairs of differently occupied participating sites. Simultaneously, 
  the total energy is increased by the repulsion terms of the 702 pairs 
  of equally occupied sites. Focusing only on the energy change related 
  to the central cube, we utilize $1/27$ of all these terms as 
  approximation of $f_{ij}$, namely the sum of the interaction of the 
  occupied site in the central cube with the empty sites in all cubes, 
  reduced by the sum of the interactions of the occupied site in the 
  central cube with the occupied sites in the neighboring cubes.
\end{thebibliography}
\end{document}